\documentclass[twocolumn]{book}
\usepackage[dvips]{graphicx,color}
\usepackage{makeidx,universe}
\usepackage{multirow}

\makeauthorindex

\BookTitle{Proceedings of the XXIX PHYSICS IN COLLISION}

\CopyRight{\copyright 2009 by Universal Academy Press, Inc.}

\begin{document} 

\pagenumbering{arabic}

\chapter{%
{\LARGE \sf
Multi-boson production
} \\
{\normalsize \bf 
Paolo Mastrandrea $^{\ast,\dagger}$ } \\
{\small \it \vspace{-.5\baselineskip}
$\ast$ Fermilab, 
      Batavia, IL 60510-5011, USA \\
$\dagger$ On behalf of CDF and D\O{} collaborations
}
}


\AuthorContents{P.\ Mastrandrea}

\AuthorIndex{Mastrandrea}{P.}

  \baselineskip=10pt 
  \parindent=10pt    

\section*{Abstract} 

The studies of the diboson production in $p \bar{p}$ collisions
at 1.96 TeV performed by CDF and D\O{} collaborations at the Tevatron
collider are reported in this paper.
The diboson events are identified by means of both leptonic and semi-leptonic
final states.
The presented results use different statistical samples collected
by the Tevatron up to 4.8 fb$^{-1}$.
Measured production cross sections are in good agreement with Standard Model
predictions and the limits on the anomalous triple gauge boson couplings are
competitive with those measured by experiments at the
Large Electron-Positron collider (LEP).

\section{Introduction} 

\begin{figure}[t]

 \begin{tabular}{cc}
  \begin{minipage}{.45\hsize}
   \begin{center}
     \includegraphics[width=.98\textwidth]{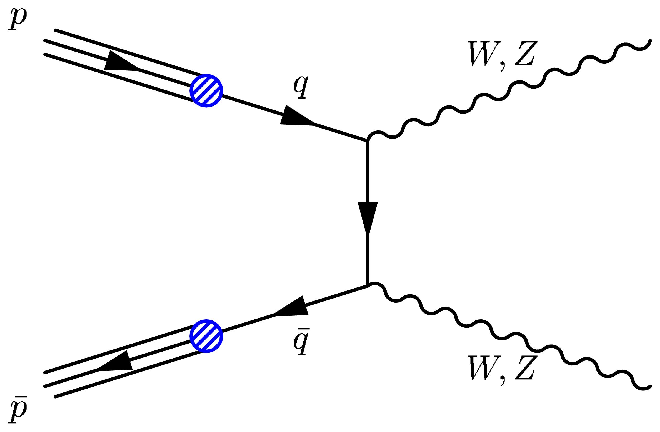}
   \end{center}
  \end{minipage}

  \begin{minipage}{.45\hsize}
   \begin{center}
     \includegraphics[width=.98\textwidth]{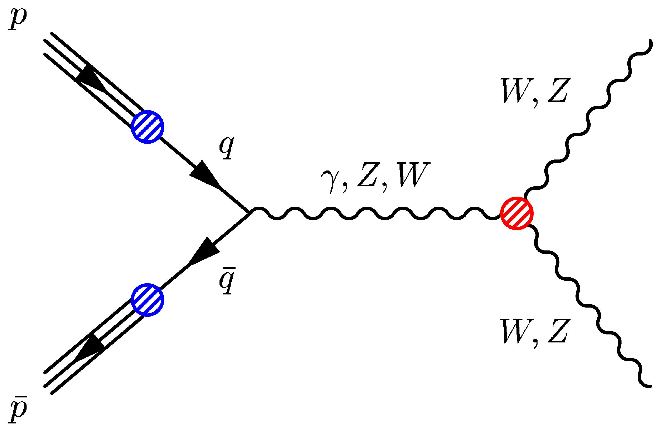}
   \end{center}
  \end{minipage}
 \end{tabular}

 \caption{Diboson production via particle-antiparticle annihilation
   ($t$-channel) on the left, and via boson self-interaction
   ($s$-channel) on the right.}

 \label{fig:multibosonproduction}
\end{figure}

Measurements of the associated production of two vector bosons ($\gamma$, $W$, $Z$)
are important tests of the electroweak sector of the Standard Model (SM).
The production of a diboson final state (Fig. \ref{fig:multibosonproduction})
can occur by particle-antiparticle annihilation ($t$-channel)
or by boson self-interaction ($s$-channel).
This kind of final state represents a unique probe for the
\emph{triple gauge boson coupling} (TGC).
Vector boson self-couplings are fundamental predictions of the SM resulting from
the non-Abelian nature of the $SU(2)_L \times U(1)_Y$ gauge symmetry theory,
which was firstly demonstrated by precision measurement of $W^+W^-$ and $ZZ$
pair production at LEP II \cite{PDG}.
Observing TGCs not permitted in the SM or with different intensity with respect to the
SM predictions, would be a sign of new physics.
A summary of the different TGC's is reported in Table \ref{tab:couplings}

\begin{table}[b]
  \small
  \caption{Summary of the triple gauge boson couplings.}
  \begin{center}
    \begin{tabular}{|c|c@{$\to$}c|c|}
      \hline
      Coupling & \multicolumn{3}{|c|}{Final state}\\
      \hline
      \hline
      \multirow{4}{*}{$VWW$ {\small $(V = Z, \gamma)$}} & $W$ & $W \gamma$ &
                                          \multirow{2}{*}{Not present at LEP}\\
                      & $W     $ & $W Z$      & \\ \cline{2-4}
                      & $\gamma$ & $W W$      & \multirow{2}{*}{LEP and Tevatron}\\
                      & $Z     $ & $W W$      & \\
      \hline
      \hline
      \multirow{2}{*}{$Z \gamma \gamma^{\ast}$ and $Z \gamma Z^{\ast}$} & $\gamma^{\ast}$ & $Z \gamma$ &
                                                   \multirow{4}{*}{Absent in SM}\\
                             & $Z^{\ast}$ & $Z \gamma$      & \\
       \cline{1-3}
      \multirow{2}{*}{$Z Z \gamma^{\ast}$ and $Z Z Z^{\ast}$} & $\gamma^{\ast}$ & $Z Z$ & \\
                             & $Z^{\ast}$ & $ Z Z$      & \\
      \hline
    \end{tabular}
  \end{center}
  \label{tab:couplings}
\end{table}

The $VWW$ vertices are completely described by 14 independent couplings,
7 each for $ZWW$ and $\gamma WW$ processes.
Assuming $C$ and $P$ conservation and electromagnetic gauge invariance, the number
of independent parameters can be reduced to 5.
A common adopted set is
$(\kappa_{\gamma}, \ \kappa_Z, \ \lambda_{\gamma}, \ \lambda_Z, \ g_1^Z)$,
where
$\kappa_{\gamma} = \kappa_Z = g_1^Z = 1$ and
$\lambda_{\gamma} = \lambda_Z = 0$,
in the SM at tree level.
Using constraints due to gauge invariance, $\kappa_Z$ and $\lambda_Z$ can be
expressed as a function of the other parameters and the weak mixing angle $\theta_W$ as:
$\kappa_Z = g_1^Z - (\kappa_{\gamma} - 1) \tan^2 (\theta_W)$ and
$\lambda_Z = \lambda_{\gamma}$.

A common formulation to report the independent parameters describing
the TGC in $VWW$ vertices is
$(\Delta \kappa_{\gamma}, \ \Delta g_1^Z, \ \lambda_{\gamma})$,
where $\Delta \kappa_{\gamma} = \kappa_{\gamma} - 1$ and
$\Delta g_1^Z = g_1^Z - 1$.

Deviations from the SM prediction for the
$Z \gamma \gamma^{\ast}$ and $Z \gamma Z^{\ast}$ couplings may be described
in terms of 4 parameters each, $h_i^V$ $(i = 1,..,4)$, respectively for $V = \gamma$
and $V = Z$.
All these anomalous contributions to the cross section increase rapidly
with the center-of-mass energy.
In order to ensure unitarity, these parameters are usually described by a
form-factor representation, $h_i^V(s) = h_{i0}^V / (1 + s / \Lambda^2)^n$,
where $\Lambda$ is the energy scale for the manifestation of a new phenomenon
and $n$ is a sufficiently large power.
Usual values adopted for $\Lambda$ in the presented results are $1.5$ TeV and $2$ TeV.

Deviations from the SM prediction for the
$Z Z \gamma^{\ast}$ and $Z Z Z^{\ast}$ couplings may be described
in terms of 2 parameters each, $f_i^V$ $(i = 4,5)$, respectively for $V = \gamma$
and $V = Z$.

All the couplings $h_i^V$ and $f_i^V$ are zero at tree level in the SM.

\section{Methods} 

All the presented analysis have been performed utilizing the data collected by
the CDF and D\O{} detectors installed at the Fermilab's Tevatron, an
accelerator capable to collide proton with antiproton at a center-of-mass
energy $\sqrt{s} = 1.96$ TeV and an instantaneous luminosity
$\mathcal{L} = 3 \cdot 10^{32}$ cm$^{-2}$ s$^{-1}$.
Both CDF and D\O{} are general purpose detectors, cylindrically symmetric
around the beam axis which is oriented as the $z$ direction.
The polar angle $\theta$ is measured from the origin of the coordinate system
at the center of the detector with respect to the $z$ axis.
The pseudorapidity, transverse energy and transverse momentum are defined as
$\eta = -\ln \tan (\theta / 2)$, $E_T = E \sin (\theta)$ and $p_T = p \sin (\theta)$,
respectively.
Detailed descriptions of the features of the two detectors can be found in
\cite{CDF_DETECTOR} and  \cite{D0_DETECTOR}.

Several combinations of the $W^{\pm}$ and $Z$ bosons decay channels are utilized to
reconstruct the diboson events.
High-$p_T$ electrons and muons are widely used in the reconstruction of the
vector bosons candidates because of the clear signatures provided
in an hadronic environment.
Neutrinos are identified evaluating the missing transverse energy ($\not \!\!\! E_T$)
in the event.
The missing $E_T$ is defined by $\not \!\!\! E_T = | \vec{\not \!\!\! E_T} |$, {}
$\vec{\not \!\!\! E_T} = - \sum_i E_T^i \mathbf{\hat{n}_i} $, {}
where $\mathbf{\hat{n}_i} $ is a unit vector perpendicular to the beam axis
and pointing at the $i^{th}$ calorimeter tower.
The index $i$ runs over all the calorimeter towers with an energy deposit
over a minimum threshold.

An anomalous TGC can affect both the production cross section and
kinematics observables of the diboson final states.
The $p_T$ or $E_T$ distributions of fully reconstructed candidate bosons
or of the charged leptons from boson decays are utilized to extract
limits on the TGC parameters.
These distributions are compared with MC predictions for different set of
coupling constants in order to evaluate $95 \%$ C.L. limits.

\section{Results} 

The CDF and D\O{} collaborations are producing many competitive measurements
in the diboson sector.
In the following the most updated results are presented.
Only peculiar features will be discussed.

\subsection{Z$\gamma$} 

The measurement of the cross section for the production of $Z \gamma$
events, with the $Z$ boson decaying in two neutrinos and the photon with
$E_T > 90$ GeV and $|\eta| < 1.1$, has been performed by the D\O{} collaboration
using a sample of data corresponding to an integrated luminosity of
$3.6$ \mbox{fb$^{-1}$} \cite{D0_Zgamma}.
The analysed events have been collected with a trigger from a set of
high-$E_T$ single EM-cluster triggers, which are ($99 \pm 1) \%$ efficient
for photons of $E_T > 90$ GeV.
The events selection requires also the presence of a single photon candidate
and $\not \!\!\! E_T > 70$ GeV.
The background to the $\gamma + \not \!\!\! E_T$ signal is mainly due to
misidentified $W$ or $Z$ events and to events in which muons from the beam halo
or cosmic rays undergo bremsstrahlung and produce energetic photons.
This non-collision background is reduced using a pointing algorithm which
utilizes the transverse and longitudinal energy distributions in the
electromagnetic (EM) calorimeter and central preshower detector, to estimate
the position of production vertex ($z_{EM}$) along the beam direction ($z$ axis).
The algorithm assumes that given EM showers are initiated by photons and utilizes
the distance of closest approach of the direction of the EM shower to
the $z$ axis.
Selected events are required to have $|z_{EM} - z_V| < 10$ cm, where
$z_V$ is the $z$ position of the reconstructed primary vertex of the event.

After applying all selection criteria 51 candidate events are observed,
with a predicted background of $17 \pm 0.6(stat.) \pm 2.3(syst.)$ events.
The measured cross section for the production of $Z \gamma$ events with
$Z \to \nu \overline{\nu}$ and the photon with $E_T > 90$ GeV is:
\[\sigma = 32 \pm 9 (stat. + syst.) \pm 2 (lumi.) \ \textnormal{fb}\]
This result is in good agreement with the next-to-leading order (NLO)
prediction of $39 \pm 4$ fb.
The statistical significance of this measurement has been estimated
performing $10^8$ background-only pseudo-experiments: the p-value is evaluated
as the fraction of pseudo-experiments with an estimated cross section above
the measured one.
This probability is found to be $3.1 \cdot 10^{-7}$, which corresponds to
a statistical significance of 5.1 standard deviations, making this the
first observation of the $Z \gamma \to \nu \overline{\nu} \gamma$ process
at the Tevatron.

The limits on the anomalous triple gauge couplings (aTGC)
are evaluated comparing the photon $E_T$ spectrum
in data with that from the sum of the background and the predicted signal
for each pair of couplings for a grid in which $h_{30}^V$ runs from -0.12 to 0.12
with a step of 0.01 and $h_{40}^V$ runs from -0.08 to 0.08 with step of 0.001.
The MC samples are generated with a leading order (LO)
$Z\gamma$ generator (corrected for the
NLO with an $E_T$-dependent $K$-factor) for the form-factor $\Lambda = 1.5$ TeV.
The resulting one-dimension limits in the neutrino channel alone are
$|h_{30}^{\gamma}| < 0.036$,
$|h_{40}^{\gamma}| < 0.0019$,
$|h_{30}^Z| < 0.035$,
$|h_{30}^Z| < 0.0019$.
To further improve the sensitivity these measurements are combined with the
limit evaluated for the $Z\gamma \to ll\gamma$ ($l = e, \ \mu$) channels
on a $1$ \mbox{fb$^{-1}$} data sample, previously analysed \cite{D0_Zgamma_llgamma}
by the D\O{} collaboration.
The combination of all three channels yields the most stringent limits
on the aTGC set at a hadron collider to date, reported in Table \ref{tab:Zgamma_D0}
The limits on the $h_{30}^Z$, $h_{30}^Z$ and $h_{40}^{\gamma}$ couplings
improve on the constraints from LEP II, and are the most restrictive to date.
\\

\begin{figure}[t]

   \begin{center}
     \includegraphics[width=.40\textwidth]{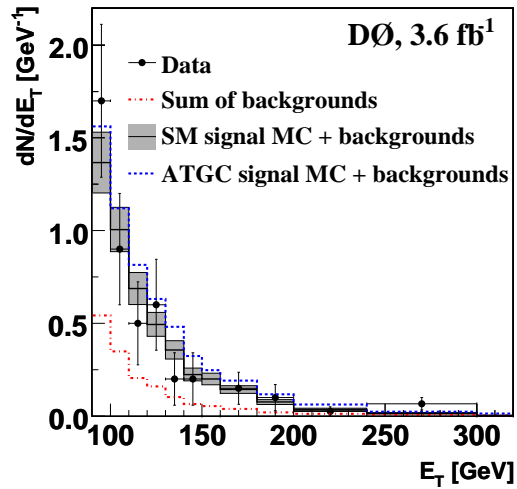}
   \end{center}

 \caption{Photon $E_T$ distribution in data, compared with the sum of backgrounds,
   the sum of MC signal and background for the SM prediction (solid line) and
   for the aTGC prediction with $h_{30}^{\gamma} = 0.09$ and $h_{40}^{\gamma} = 0.005$
   (dashed line).
   The shaded band corresponds to the $\pm 1$ standard deviation total uncertainty on the
   predicted sum of SM signal and background.}

 \label{fig:Zgamma_CDF_pictures}
\end{figure}

\begin{figure}[b]

 \begin{tabular}{cc}
  \begin{minipage}{.45\hsize}
   \begin{center}
     \includegraphics[width=.98\textwidth]{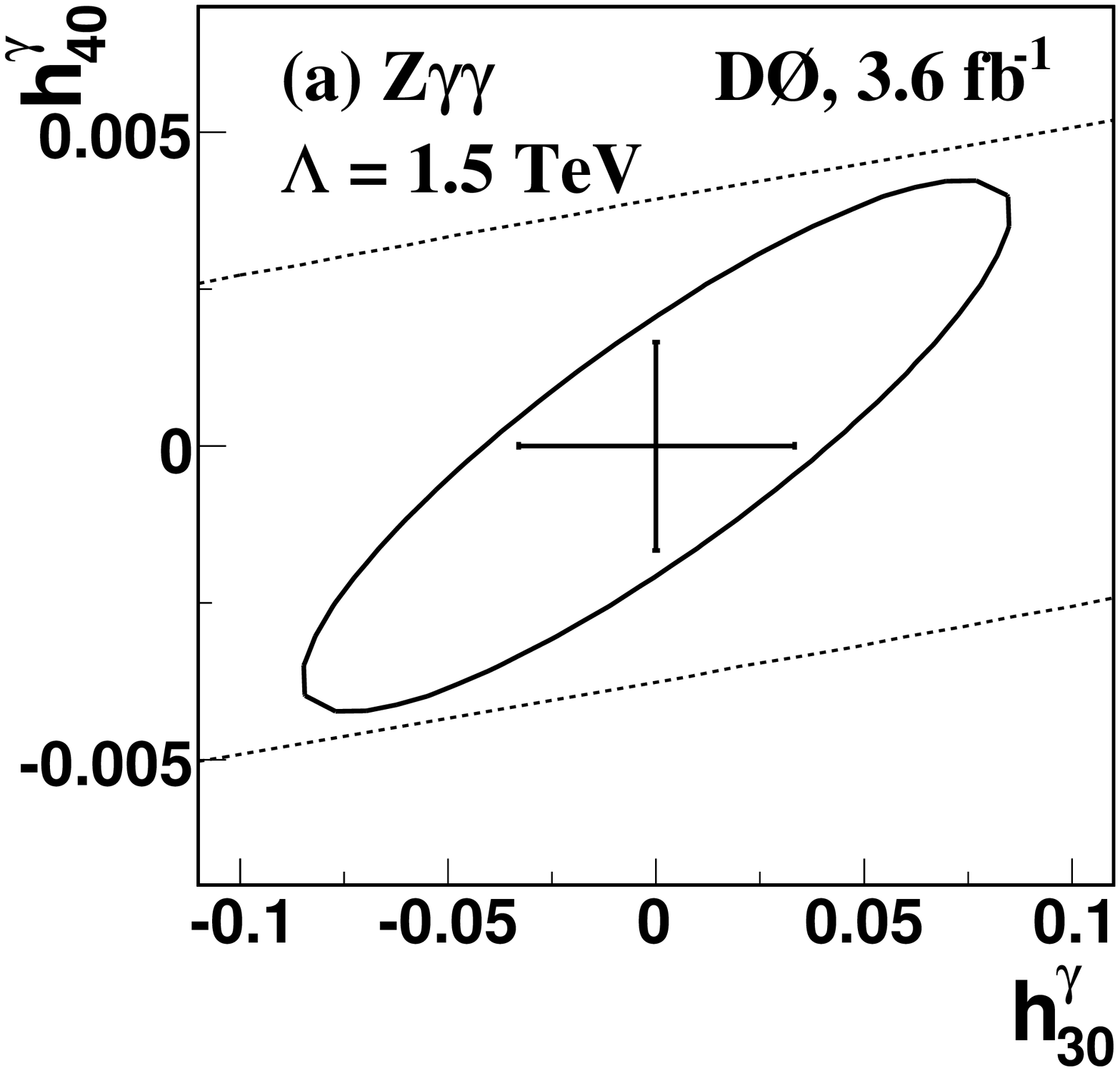}
   \end{center}
  \end{minipage}

  \begin{minipage}{.45\hsize}
   \begin{center}
     \includegraphics[width=.98\textwidth]{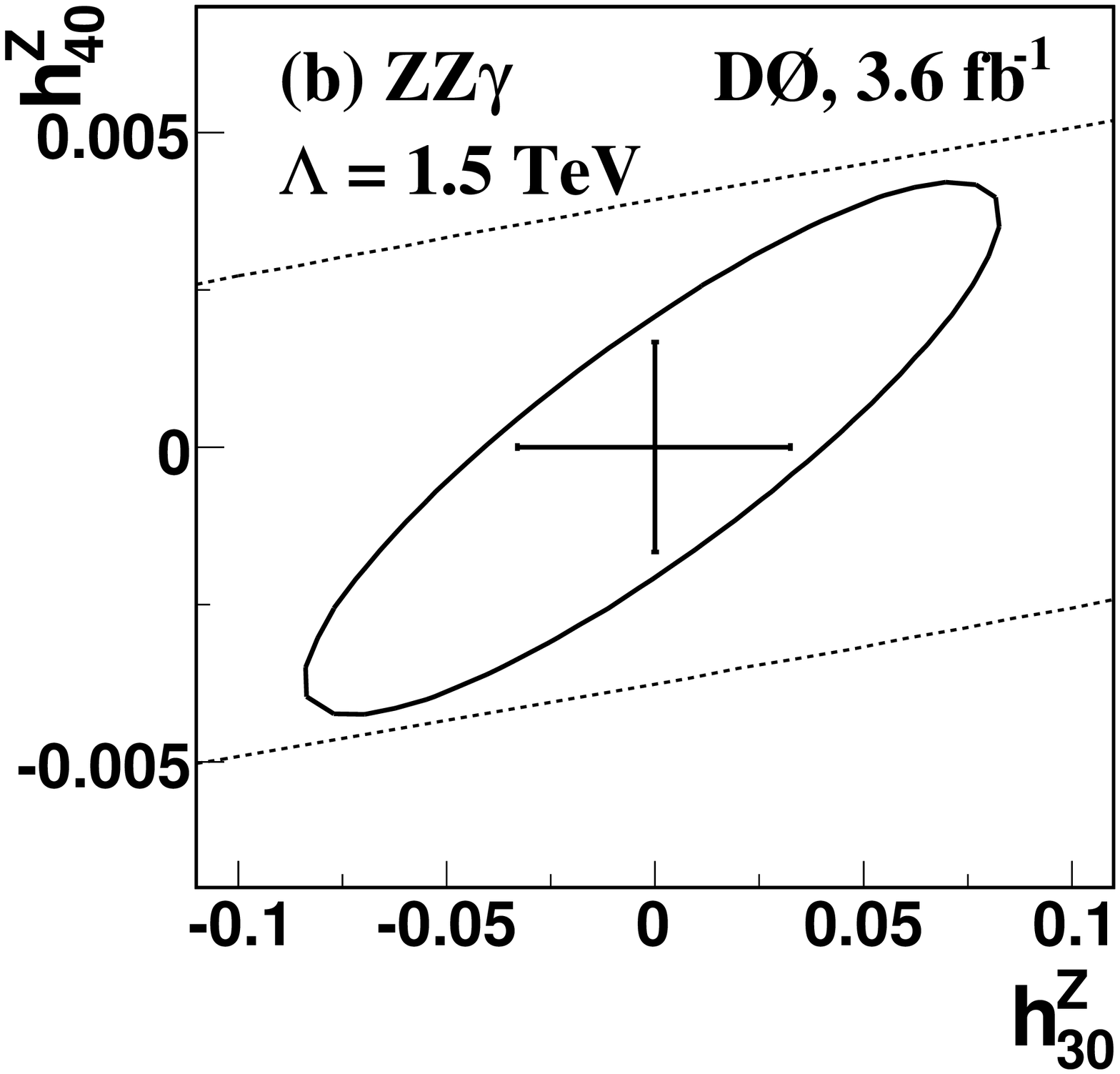}
   \end{center}
  \end{minipage}
 \end{tabular}

 \caption{Two-dimensional bounds (ellipses) at $95 \%$ C.L. on
   CP-conserving $Z\gamma\gamma$ (left) and $ZZ\gamma$ (right) couplings.
   The crosses represent the one-dimension bounds at the $95 \%$ C.L. setting
   all other couplings to zero.
   The dashed lines indicate the unitarity limits for $\Lambda = 1.5$ TeV.}

 \label{fig:Zgamma_D0_pictures}
\end{figure}

\begin{table}[!h]
  \small
  \caption{One-dimension $95 \%$ C.L. limits obtained by the D\O{} collaboration
    in the analysis of the $ZZ\gamma$ and $Z\gamma\gamma$ couplings.} 
  \begin{center}
    \begin{tabular}{|c|c|}
      \hline
      Parameter & $95\% \ C.L.$ \\
      \hline
      $|h_{30}^{\gamma}|$ & $(-0.033, +0.033)$ \\
      $|h_{40}^{\gamma}|$ & $(-0.0017, +0.0017)$ \\
      $|h_{30}^{Z}|$      & $(-0.033, +0.033)$ \\
      $|h_{40}^{Z}|$      & $(-0.0017, +0.0017)$ \\
      \hline
    \end{tabular}
  \end{center}
  \label{tab:Zgamma_D0}
\end{table}

\subsection{W$\gamma$} 

The measurement of the cross section for the associated production
of $W^{\pm}$ and $\gamma$ bosons has been performed by the CDF collaboration
on a data sample corresponding to $1$ \mbox{fb$^{-1}$} \cite{CDF_Wgamma_production}.

\begin{figure}[t]

  \begin{minipage}{.98\hsize}
   \begin{center}
     \includegraphics[width=.8\textwidth]{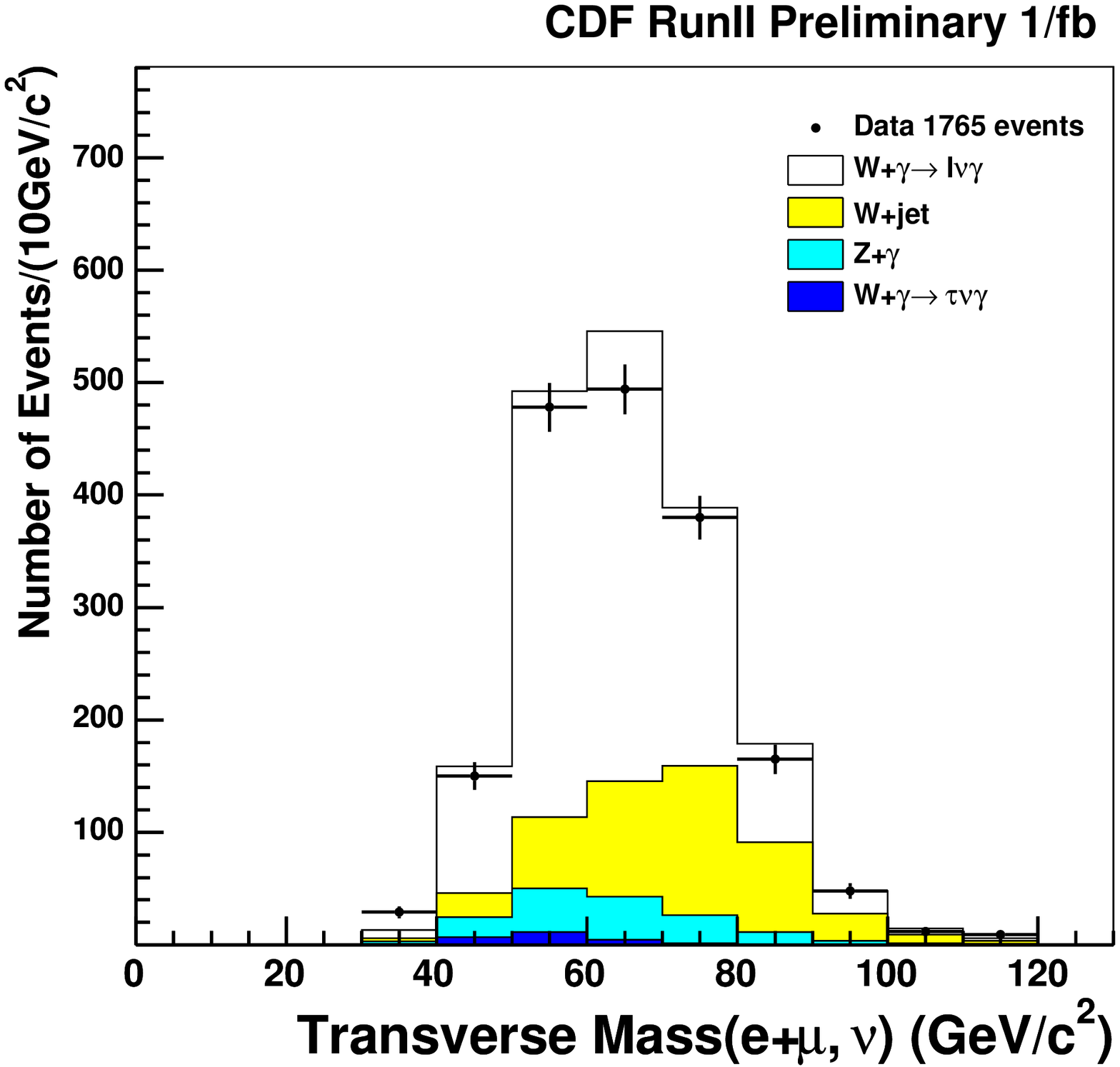}
     \includegraphics[width=.8\textwidth]{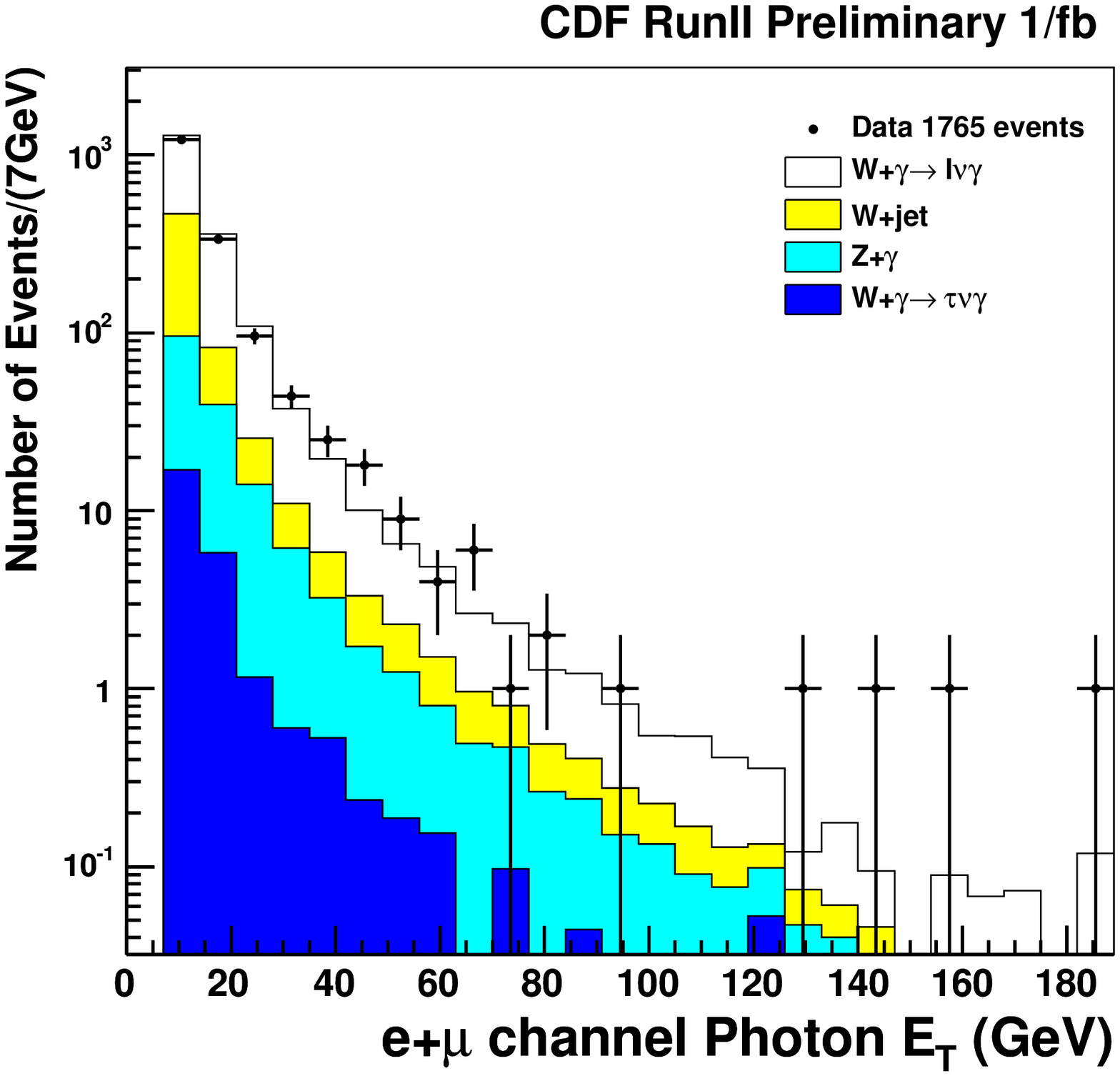}
   \end{center}
  \end{minipage}

 \caption{Transverse mass distribution of the reconstructed $W^{\pm}$ boson
   (upper), and transverse energy distribution for the candidate photon (lower).}

 \label{fig:Wgamma_CDF_pictures}
\end{figure}

The $W$ boson is reconstructed utilizing its decay in electron and neutrino
or muon and neutrino.
The \mbox{$W \to e \nu$} events are selected requiring a candidate electron with
$E_T > 25$ GeV and $|\eta| < 1.1$, and $\not \!\!\! E_T > 25$ GeV,
while the $W \to \mu \nu$ events are selected requiring a candidate muon with
$p_T > 20$ GeV/c and $|\eta| < 1.1$, and $\not \!\!\! E_T > 20$ GeV.
Furthermore the transverse mass ($M_T$) of the candidate $W$ boson is
required in the range $30 < M_T < 120$ \mbox{GeV/c$^2$}.
The candidate photons must have $E_T > 7$ GeV and $|\eta| < 1.1$.
The distribution of the transverse mass of the candidate $W$ boson and
of the transverse energy of the candidate photon are compared in
Fig. \ref{fig:Wgamma_CDF_pictures} with the simulation expectation.

The measured cross section for the production of $W \gamma$ events, with the
$W$ boson decaying in $e \nu$ or $\mu \nu$ and the photon with $E_T > 7$ GeV is:
\[ \sigma = 18.03 \pm 0.65 (stat.) \pm 2.55 (sys) \pm 1.05 (lumi.) \ \textnormal{pb}. \]
This result is in good agreement with the SM prediction of $19.3 \pm 1.4$ pb.

\begin{table}[b]
  \small
  \caption{One-dimension $95 \%$ C.L. limits obtained by the D\O{} collaboration
    in the analysis of the $W\gamma$ production.} 
  \begin{center}
    \begin{tabular}{|c|c|}
      \hline
      Parameter & $95\% \ C.L.$ \\
      \hline
      $\Delta \kappa_{\gamma}$ & $(-0.88, +0.96)$ \\
      $\lambda_{\gamma}$       & $(-0.20, +0.20)$ \\
      \hline
    \end{tabular}
  \end{center}
  \label{tab:Wgamma_D0}
\end{table}

\begin{figure}[t]

  \begin{minipage}{.88\hsize}
   \begin{center}
     \includegraphics[width=.8\textwidth]{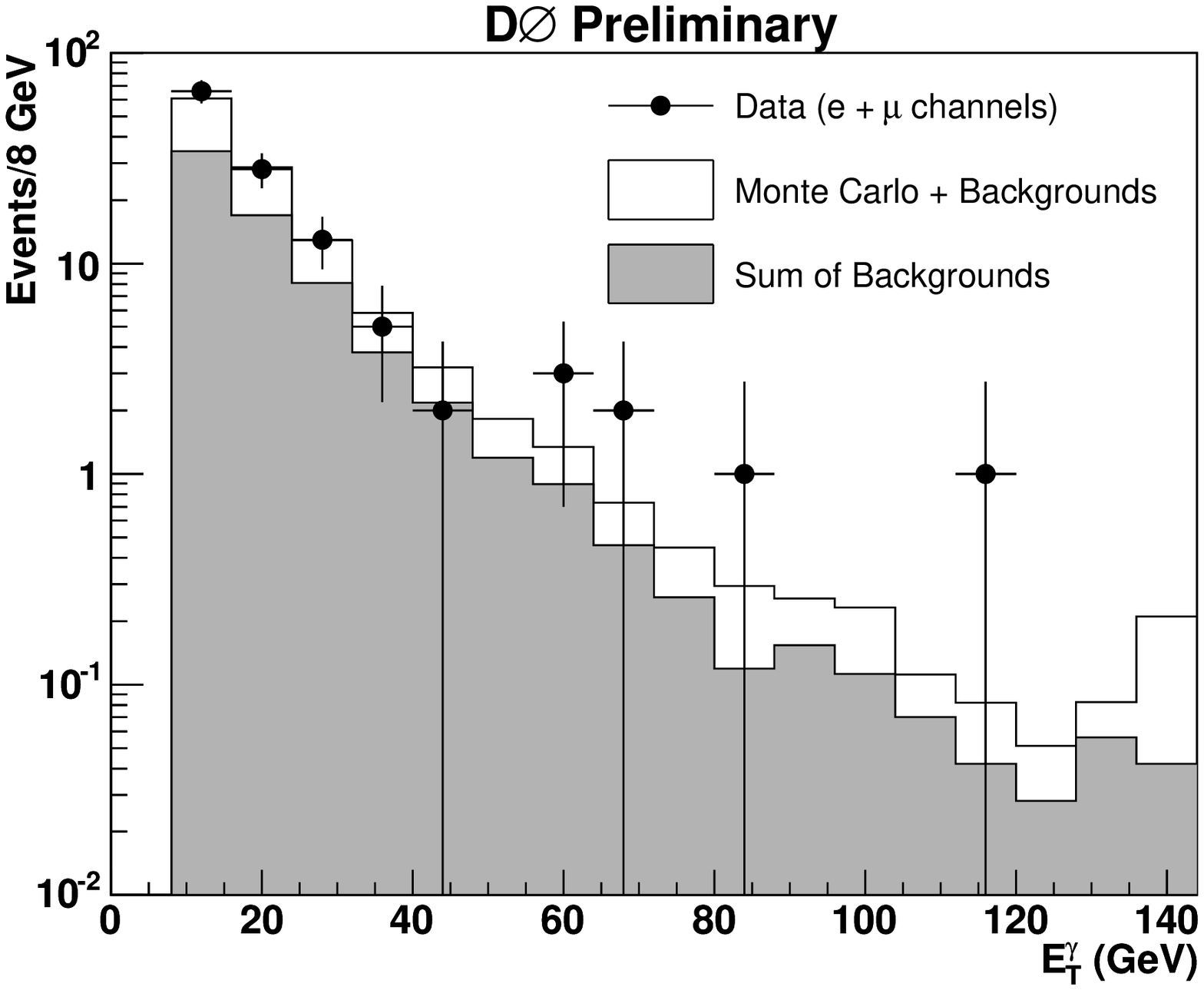}
     \includegraphics[width=.8\textwidth]{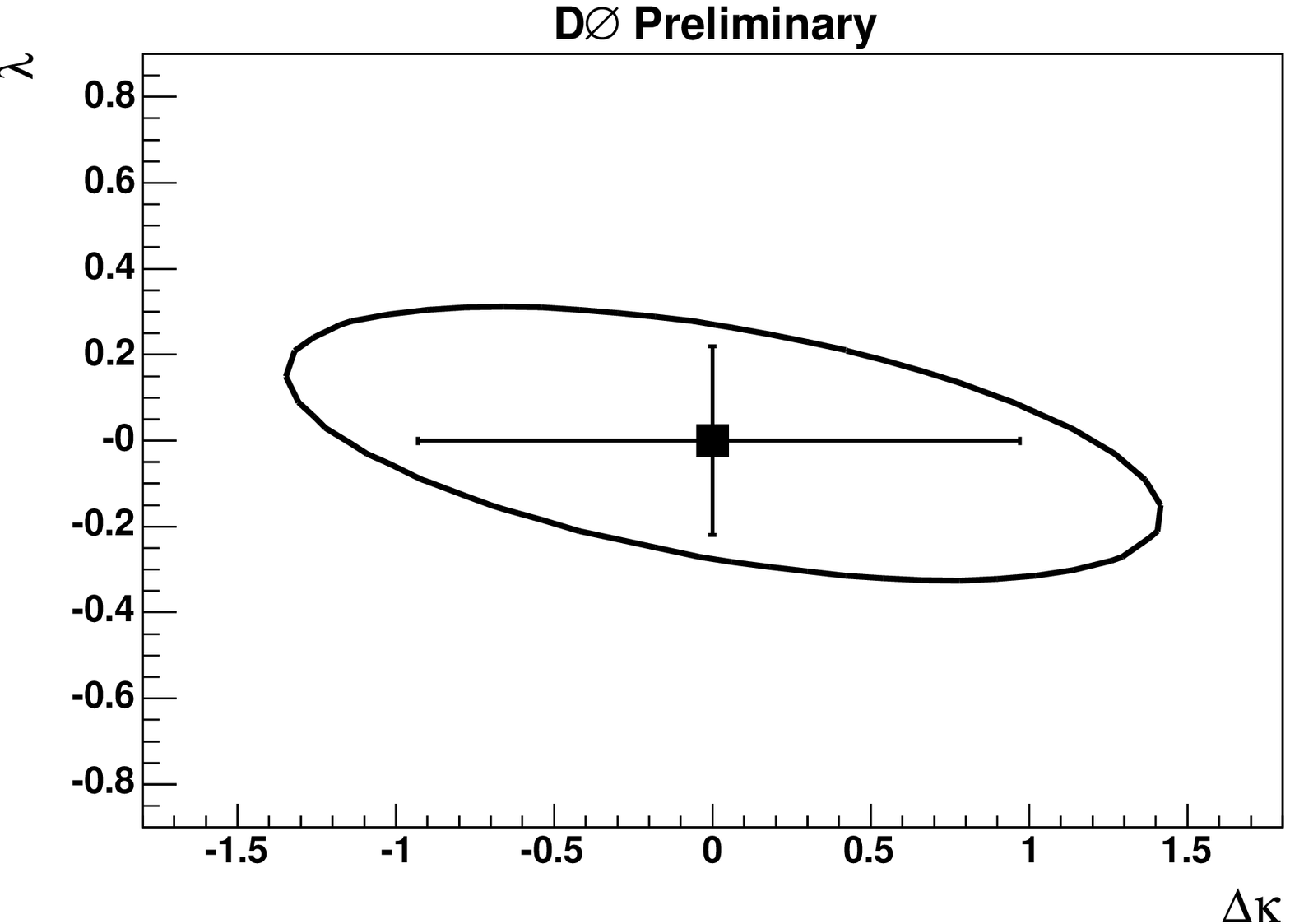}
   \end{center}
  \end{minipage}

 \caption{Transverse energy distribution of the reconstructed photon
   (upper), and comparison between one-dimension and two-dimension limit
   at $95 \%$ C.L. on the aTGC paramenters for the $WW\gamma$ coupling (lower).}

 \label{fig:Wgamma_D0_pictures}
\end{figure}

An analysis using analogous techniques and performed by the D\O{} collaboration
on a data sample corresponding to $162$ \mbox{pb$^{-1}$} \cite{D0_Wgamma_aTGC}
evaluated $95 \%$ C.L. limits on $WW\gamma$ aTGC.
The limits are extracted from the comparison of the $E_T$ distribution of
the candidate photon with the theoretical predictions as a function of the
coupling parameters.
The measured limits at $95 \%$ C.L. are reported in Table \ref{tab:Wgamma_D0},
and the comparison with the two-dimensional limit 
is reported in Fig. \ref{fig:Wgamma_D0_pictures}

\subsection{WW} 

The measurement of the production cross section of $WW$ events
in $p \overline{p}$ collisions has been performed
by the CDF collaboration on a data sample corresponding to $3.6$ \mbox{fb$^{-1}$}
\cite{CDF_WW_production}.

The events are reconstructed using the decay chain
$W^+W^- \to l^+ \nu l^- \overline{\nu}$, where $l = e, \ \mu, \ \tau$,
with final states $e^+e^-$, $\mu^+\mu^-$ and $e^{\pm}\mu^{\mp}$,
and acquired with high-$p_T$ electron or muon triggers.
Several processes contribute to the background:
Drell-Yan where the measured large $\not \!\!\! E_T$ is due to resolution tails,
$WZ \to lll\nu$ where one lepton is lost,
$t \overline{t} \to b \overline{b}ll\nu\nu$, and
$W\gamma$ and $W+$jets where a photon or jet is misidentified as a lepton.
The majority of $WW$ events have zero jets, and as the jet multiplicity increases
the backgrounds change (e.g. $t \overline{t}$ becomes a larger
contribution with increasing jet multiplicity).
Events with reconstructed jets with $E_T > 15$ GeV and $|\eta| < 2.5$ are
rejected.
To suppress the Drell-Yan background, dielectron and dimuon events are required
to have $\not \!\!\! E_{Tspec} > 25$ GeV and $\not \!\!\! E_{Tspec} > 15$ GeV
for electron-muon events, where $\not \!\!\! E_{Tspec}$ is defined as:
\[\not \!\!\! E_{Tspec} \equiv \left\{
\begin{array}{ll}
\not \!\!\! E_T & if \ \Delta\phi(\not \!\!\! E_T, l, j) > \frac{\pi}{2}\\
\not \!\!\! E_T \sin(\Delta\phi(\not \!\!\! E_T, l, j) )  &
if \ \Delta\phi(\not \!\!\! E_T, l, j) < \frac{\pi}{2}
\end{array} \right.
\] 
\begin{figure}[t]
  \begin{minipage}{\hsize}
    \begin{center}
      \includegraphics[width=.98\textwidth]{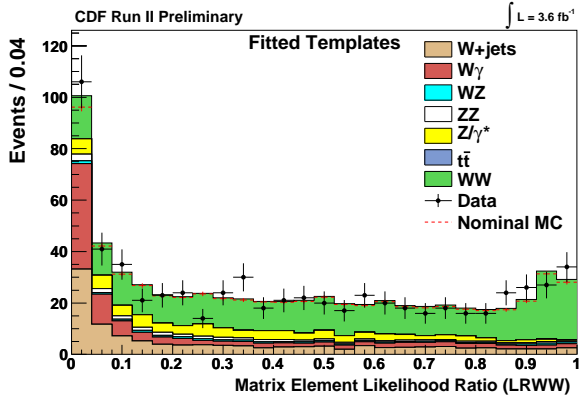}
    \end{center}
    \end{minipage}
 \caption{Likelihood ratio distribution for data compared with
   fitted signal and backgrounds templates and with the nominal MC prediction.}

 \label{fig:WW_CDF_picture}
\end{figure}
This definition is a requirement that $\vec{\not \!\!\! E_T}$ transverse to each lepton
or jet in the event is greater than the minimum threshold if $\vec{\not \!\!\! E_T}$
points along that object, so that losing energy from just one such object in an event
would not allow it to enter the sample.
Furthermore, in order to suppress the heavy flavor contribution, the invariant mass
of the two selected leptons is required to be $M_{l^+l^-} > 16$ \mbox{GeV/c$^2$}.

To measure the $WW$ production cross section, a matrix element probability for each event
to be a $WW$ event has been calculated.
This probability is used to build a likelihood ratio discriminant, and the normalization
of $WW$ process is extracted by a fit, using MC templates.

The event-by-event matrix element probability density $P_m(x_{obs})$ is defined for
the four processes ($m$) $WW$, $ZZ$, $W\gamma$ and $W+$jets, as:
\begin{equation}
  P_m(x_{obs}) = \frac{1}{<\sigma_m>} \int \frac{d \sigma_m^{th}(y)}{dy}
  \epsilon(y)G(x_{obs},y) dy
  \label{eq:WW_CDF_pdf_me}
\end{equation}
where

\begin{flushleft}
\begin{tabular}{lp{0.37\textwidth}}
$x_{obs}$              & are the observed ``leptons'' and $\not \!\!\! E_T$, \\
$y$                    & are the true lepton four-vectors (include neutrinos), \\
$\sigma_m^{th}$        & is leading-order theoretical calculation of the cross-section
 for the process $m$,\\
$\epsilon(y)$          & is total event efficiency $\times$ acceptance,\\
$G(x_{obs},y)$         & is an analytic model of resolution effects,\\
$\frac{1}{<\sigma_m>}$ & is the normalization.
\end{tabular}
\end{flushleft}

Since the neutrinos are not reconstructed, the unobserved degrees of freedom (DOF)
are integrated out in equation (\ref{eq:WW_CDF_pdf_me}), reducing the number of the DOF
to eight, measured in the selected events.

The event probability densities are used to construct a likelihood ratio discriminant:
\[
LR_{WW}(x_{obs}) \equiv \frac{P_{WW}(x_{obs})}{P_{WW}(x_{obs}) + \sum_i k_i P_i(x_{obs})}
\]
where $k_i$ is the expected fraction for each background, with $\sum_i k_i = 1$.

A binned maximum likelihood method is used to extract the $WW$ cross section using
the shape of the $LR_{WW}$ distributions for signal and backgrounds along with their
estimated normalizations and systematic uncertainties.
The likelihood function is formed from a product of Poisson probabilities for each
bin in $LR_{WW}$ distribution (Fig. \ref{fig:WW_CDF_picture}).
Additionally Gaussian constraints are applied for each systematic indetermination
$S_C$ \footnote{Details about the systematic uncertainties can be found in
\cite{CDF_WW_production}}.
The likelihood function is defined as:
\begin{equation}
\mathcal{L} = \left(\prod_i \frac{\mu_i^{n_i} e^{-\mu_i}}{n_i!} \right)
\cdot \prod_C e^{\frac{S_C^2}{2}}
\end{equation}
where $n_i$ is the number of data events in the $i$-th bin and
$\mu_i$ is the total expectation for the $i$-th bin, given by:
\begin{equation}
\mu_i = \sum_k \alpha_k \left[ \prod_C (1 + f_k^C S_C) \right] (N_k^{Exp})_i
\label{eq:CDF_WW_expectedvalue}
\end{equation}
This formulation includes the proper correlation of the different systematic
uncertainties.
In equation (\ref{eq:CDF_WW_expectedvalue})
$(N_k^{Exp})_i$ and $f_k^C$ are the expected number of events in the $i$-th bin
and the fractional uncertainty associated with the systematic $S_C$
for the process $k$.
The parameter $\alpha_k$ is the ratio between the measured cross section
and the predicted one: it is freely floating
for the $WW$ process and is fixed to 1 for all other processes.

The measured cross section for the production of $WW$ events is:
\[
\sigma(p \overline{p} \to WW) = 12.1 \pm 0.9 \ (stat.) \ {}_{-1.4}^{+1.6} \ (syst.) \ \textnormal{pb}
\]
This is in good agreement with the theoretical prediction
$\sigma(p \overline{p} \to WW) = 11.66 \pm 0.70$ pb
and is the best measurement to date of the production cross section
of the $p \overline{p} \to WW$ process.

The $p_T$ distribution of the leading lepton for each selected event is used to
extract $95 \%$ C.L. limits on the $ZWW$ and $\gamma WW$ aTGC parameters.
The measured limits are reported in Table \ref{tab:WW_aTGC}

The D\O{} collaboration analysed the $WW$ production \cite{D0_WW_aTGC}
in a data sample corresponding
to $1$ \mbox{fb$^{-1}$}, utilizing the final states described above, with
competitive results for both the production cross section measurement
and the aTGC limits evaluation.
The measured cross section:
\[
\sigma(p \overline{p} \to WW) = 11.5 \pm 2.1 \ (stat.+syst.) \pm 0.7 \ (lumi.) \ \textnormal{pb}
\]
is consistent with the SM expectation of $12.0 \pm 0.7$ pb.
To enhance the sensitivity to anomalous couplings selected events are sorted
according to both leading and trailing lepton $p_T$ into a two-dimensional histogram.
For each bin the selected number of $WW$ events produced is parametrized by a quadratic
function in three-dimensional
($\Delta \kappa_{\gamma}, \lambda_{\gamma}, \Delta g_1^Z$)
space or two-dimensional
($\Delta \kappa, \lambda$)
space, as appropriate for the TGC relationship scenario under study.
In the three-dimensional case, coupling parameters are investigated in pairs,
with the third parameter fixed to the SM value.
A likelihood surface is generated by considering all channels simultaneously,
integrating over the signal,background, and luminosity uncertainties
with Gaussian distributions using the methodology described in \cite{D0_WW_aTGC}.
The measured one-dimension $95 \%$ C.L. limits for $\Lambda = 2$ TeV
are reported in Table \ref{tab:WW_aTGC}

\begin{figure}[t]

 \begin{tabular}{cc}
  \begin{minipage}{.45\hsize}
   \begin{center}
     \includegraphics[width=.98\textwidth]{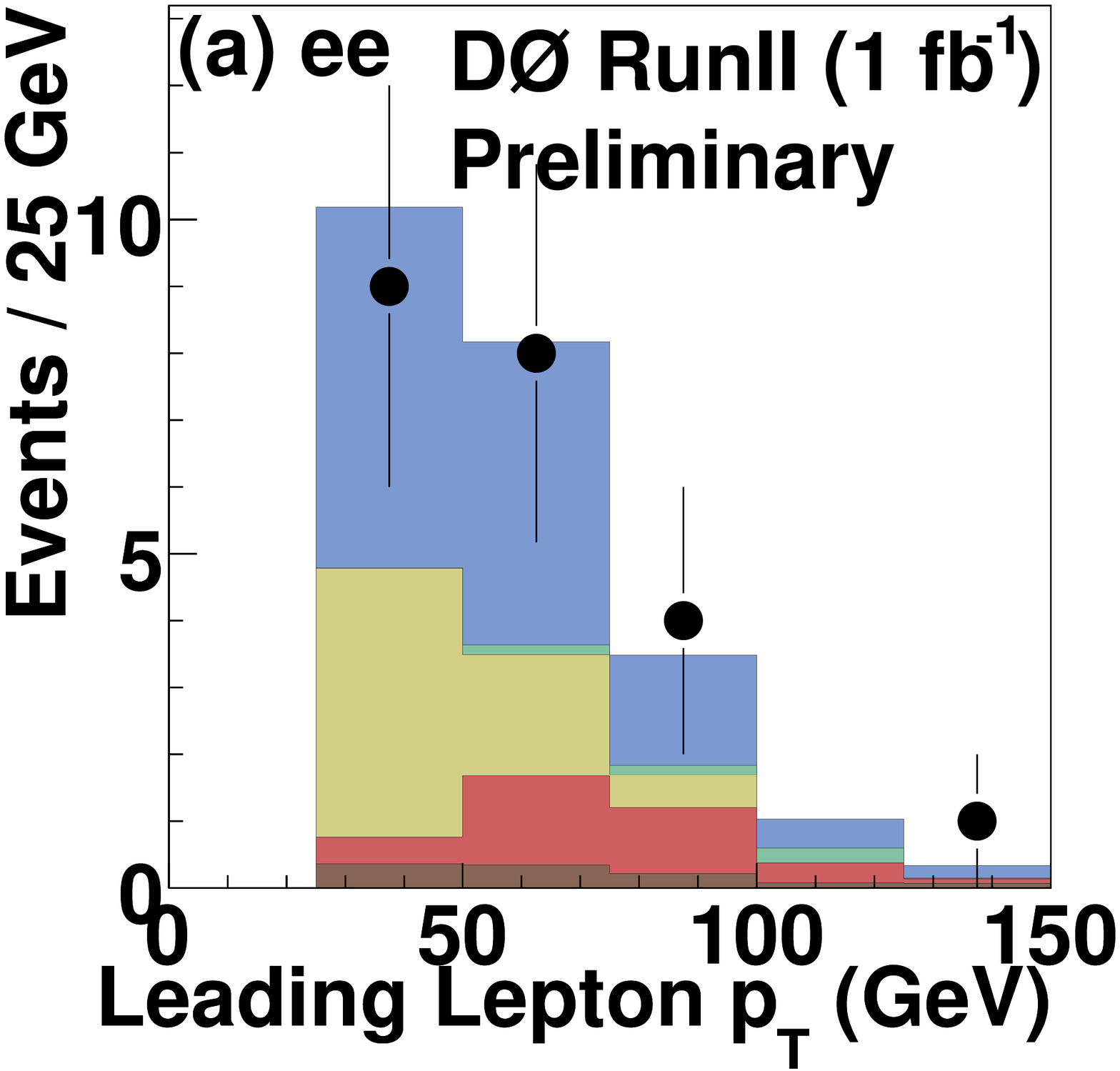}
   \end{center}
  \end{minipage}

  \begin{minipage}{.45\hsize}
   \begin{center}
     \includegraphics[width=.98\textwidth]{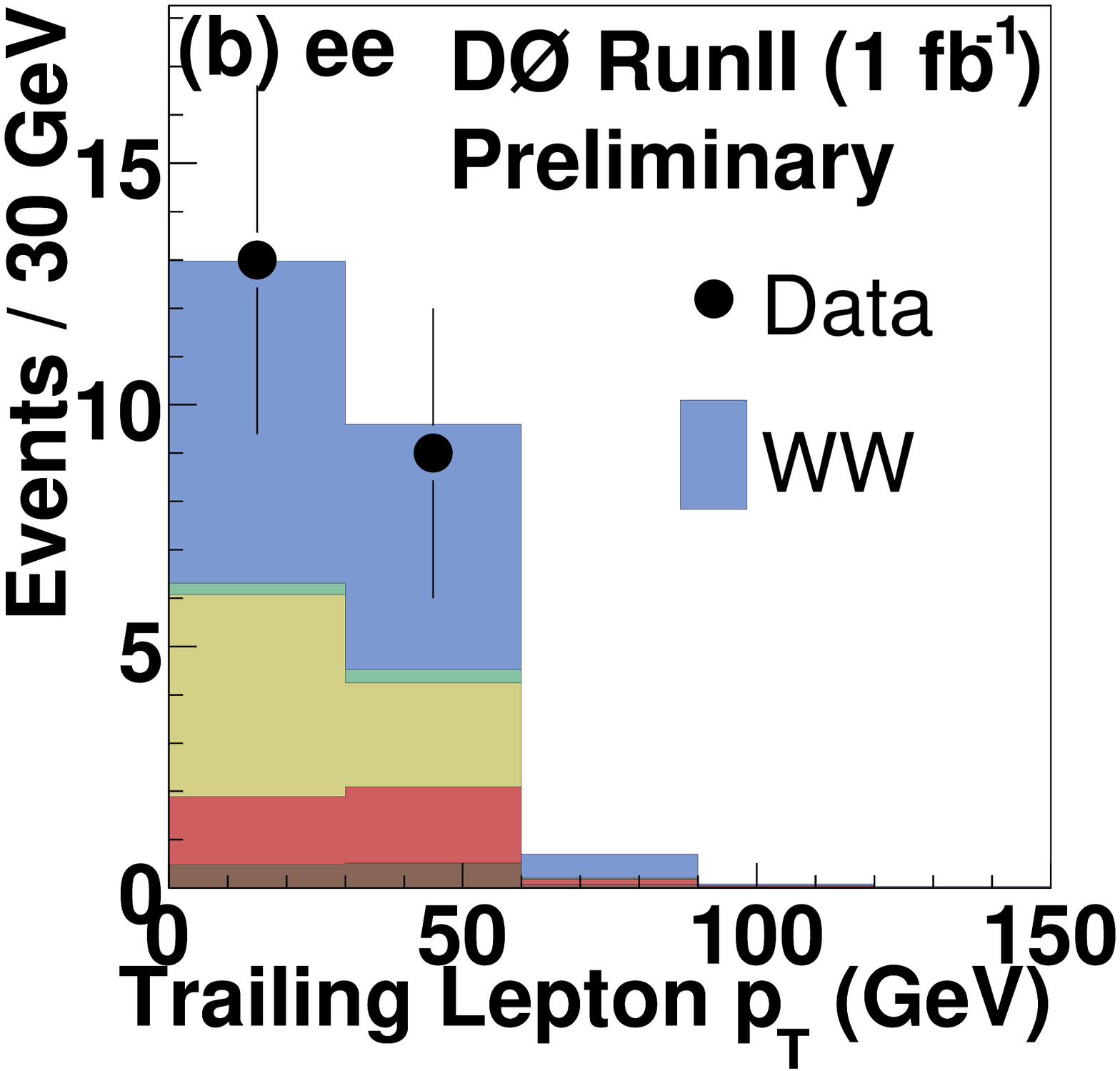}
   \end{center}
  \end{minipage}
 \end{tabular}

 \caption{Transverse momentum distribution of the leading (left) and
   trailing (right) lepton for the candidate $WW$ events, combined for all
   channels ($e^+e^- + \mu^+\mu^- + e^{\pm}\mu^{\mp}$).
   Data are compared to the estimated signal and background sum, using
   $\sigma(p \overline{p} \ to WW) = 12$ pb.}

 \label{fig:WW_D0_pictures}
\end{figure}

\begin{table}[!h]
  \small
  \caption{One-dimension $95 \%$ C.L. limits for $\Lambda = 2$ TeV obtained
    by CDF and D\O{} collaborations in the analysis of the $WW$ production.} 
  \begin{center}
    \begin{tabular}{|c|c|c|}
      \hline
      Parameter & CDF & D\O{} \\
      \hline
      $\Delta \kappa_{\gamma}$ & $(-0.57, +0.65)$  & $(-0.54, +0.83)$ \\
      $\lambda_{\gamma}$       & $(-0.14, +0.15)$  & $(-0.14, +0.18)$ \\
      $\Delta g_1^Z$           & $(-0.22, +0.30)$  & $(-0.14, +0.30)$ \\
      \hline
    \end{tabular}
  \end{center}
  \label{tab:WW_aTGC}
\end{table}

\subsection{WZ} 

\begin{table}[!b]
  \small
  \caption{Summary of signal and background events expected for
    $WZ \to l'\nu ll$ process, in a $1.9$ \mbox{fb$^{-1}$} data sample,
     analysed by CDF collaboration.}
  \begin{center}
    \begin{tabular}{l c @{$\ \pm\ $} c @{$\ \pm\ $} c @{$\ \pm\ $} c}
      \hline
      Source               & Expected & Stat & Syst & Lumi \\
      \hline
      $Z$+jets     &   2.45 & 0.48 & 0.48 & 0.00  \\
      $ZZ$         &   1.53 & 0.01 & 0.16 & 0.09  \\
      $Z\gamma$    &   1.03 & 0.06 & 0.35 & 0.06  \\
      $t\bar{t}$   &   0.17 & 0.01 & 0.03 & 0.01  \\
      $WZ$         &  16.45 & 0.03 & 1.74 & 0.99  \\
      \hline
      Total        &  21.63 & 0.48 & 2.25 & 1.15  \\
      \hline
      Observed     &  \multicolumn{4}{c}{25}\\
      \hline
    \end{tabular}
  \end{center}
  \label{tab:WZ_CDF_events}
\end{table}

The measurement of the production cross section of $WZ$ events
in $p \overline{p}$ collisions has been performed
by the CDF collaboration on a data sample corresponding to $1.9$ \mbox{fb$^{-1}$}
\cite{CDF_WZ_3l_production}.

The events are reconstructed in final states with three charged leptons
$WZ \to l'\nu ll$, where $l = e, \mu$,
and acquired with high-$p_T$ electron or muon triggers.

\begin{figure}[t]

  \begin{minipage}{.98\hsize}
   \begin{center}
     \includegraphics[width=.7\textwidth]{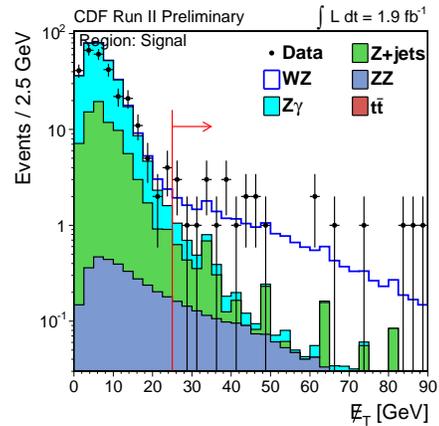}
     \includegraphics[width=.7\textwidth]{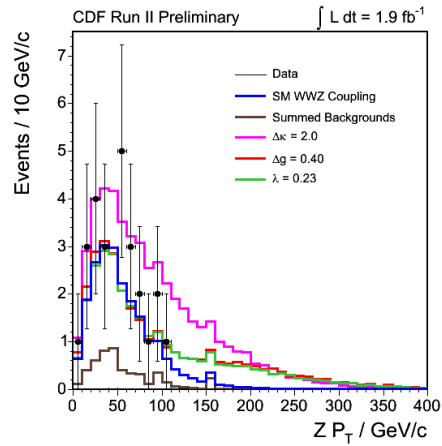}
   \end{center}
  \end{minipage}

  \caption{$\not \!\!\! E_T$ distribution of the candidates $ZW$ events (upper);
    the red line indicates the signal region.
    In the lower plot is reported the $p_T$ distribution of the candidate $Z$;
    the selected data events are compared with the MC expectation for SM TGC,
    and for three aTGC hypothesis.}

 \label{fig:WZ_3l_CDF}
\end{figure}

The $WZ$ candidates are selected from events with three identified leptons.
In order to enhance the signal acceptance, looser selection criteria are
adopted for non-leading leptons.
Aside from $WZ$ production, other SM processes that can lead to three high-$p_T$
leptons include dileptons from Drell-Yan $Z/\gamma^{\ast}$ process with an additional
lepton from a photon conversion ($Z\gamma$) or misidentified jet ($Z+$jets),
$ZZ$ where only three leptons are identified, and a small contribution from
$t \overline{t}$ production.
Aside from $t \overline{t}$, these backgrounds are strongly suppressed by requiring
$\not \!\!\! E_T > 25$ GeV in the event, consistent with the unobserved neutrino
from leptonic decay of a $W$ boson.
Furthermore the azimuthal angle between the $\not \!\!\! E_T$ direction and any
identified jet with $E_T > 15$ GeV or $WZ$ candidate lepton, is required to be
greater than $9^{\circ}$ to suppress Drell-Yan backgrounds in which the observed
$\not \!\!\! E_T$ is due to mis-measured leptons or jets.
At least one same-flavor, opposite-charge lepton pair is required in the event
with an invariant mass $M_{l^+l^-}$ in the range $[76, 106]$ \mbox{GeV/c$^2$} consistent
with leptonic $Z$ boson decay.
If more than such pair is identified in the event, the one with $M_{l^+l^-}$ closest
to the nominal $Z$ mass, is selected as $Z$ boson decay candidate pair.
In order to suppress the background from $ZZ$ production, the event is rejected if
the invariant mass of the lepton not selected in the $Z$ candidate combined with any
additional track with $p_T > 8$ GeV/c is in the range $[76, 106]$ \mbox{GeV/c$^2$}.

The $\not \!\!\! E_T$ distribution for the $ZW$ candidates is shown in
Fig. \ref{fig:WZ_3l_CDF}, while in Table \ref{tab:WZ_CDF_events}
is reported the comparison between the number of observed and expected events.
The measured cross section for the production of $ZZ$ events is:
\[
\sigma(p \overline{p} \to WZ) = 4.3^{+1.3}_{-1.0} \ (stat.) \pm 0.4 \ (syst.+lumi.) \ \textnormal{pb}
\]
in agreement with the NLO prediction
\mbox{$\sigma^{NLO}(p \overline{p} \to WZ) = 3.7 \pm 0.3$ pb}.

The $p_T$ distribution of the candidate $Z$ for each selected event
(Fig. \ref{fig:WZ_3l_CDF}) is used to
extract $95 \%$ C.L. limits on the $WWZ$ aTGC parameters.
The measured limits are reported in Table \ref{tab:ZW_CDF_aTGC}

\begin{table}[!h]
  \small
  \caption{One-dimension $95 \%$ C.L. limits for obtained
    by the CDF collaboration in the analysis of the $ZW$ production.} 
  \begin{center}
    \begin{tabular}{|c|c|c|}
      \hline
      Parameter & $\Lambda = 1.5$ TeV & $\Lambda = 2.0$ TeV \\
      \hline
      $\Delta \kappa_{Z}$      & $(-0.81, +1.29)$  & $(-0.76, +1.18)$ \\
      $\lambda_{Z}$            & $(-0.14, +0.15)$  & $(-0.13, +0.14)$ \\
      $\Delta g_1^Z$           & $(-0.14, +0.25)$  & $(-0.13, +0.23)$ \\
      \hline
    \end{tabular}
  \end{center}
  \label{tab:ZW_CDF_aTGC}
\end{table}

\subsection{ZZ} 

\begin{figure}[t]

  \begin{minipage}{\hsize}
   \begin{center}
     \includegraphics[width=.90\textwidth]{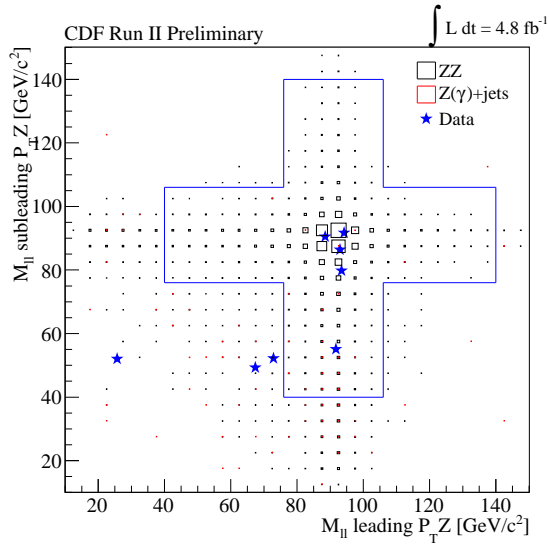}
   \end{center}
  \end{minipage}

  \caption{Distribution of the selected events in the signal region.
    The invariant mass of the two selected couples of leptons is required to be
    in ($76-106$) $GeV/c^2$, for the $Z$ candidate closest to the nominal $Z$ mass,
    and in ($40-140$) $GeV/c^2$ for the other $Z$ candidate.}

 \label{fig:ZZ_CDF_signal_region}
\end{figure}

The measurement of the production cross section of $ZZ$ events
in $p \overline{p}$ collisions has been performed
by the CDF collaboration on a data sample corresponding to $4.8$ \mbox{fb$^{-1}$}
\cite{CDF_ZZ_production}.

\begin{table}[b]
  \small
  \caption{Summary of signal and background events expected for
    $ZZ \to l_a^+l_a^-l_b^+l_b^-$ process, in a $4.8$ \mbox{fb$^{-1}$} data sample,
     analysed by CDF collaboration.}
  \begin{center}
    \begin{tabular}{cc@{$\ \pm\ $}c@{$\ \pm\ $}c}
      \hline
      Signal            & 4.68  & 0.02 (stat.)  & 0.76 (syst.)\\
      $Z(\gamma)+$jets  & 0.041 & 0.016 (stat.) & 0.029 (syst.)\\
      \hline
      Total expected    & 4.72  & 0.03 (stat.)  & 0.76 (syst.)\\
      \hline
      Observed          & \multicolumn{3}{c}{5}\\
      \hline
    \end{tabular}
  \end{center}
  \label{tab:ZZ_CDF_events}
\end{table}

The events are reconstructed in final states with four charged leptons
$ZZ \to l_a^+l_a^-l_b^+l_b^-$, where $l_a, l_b = e, \mu$,
and acquired with high-$p_T$ electron or muon triggers.
The main source of background is given by events $Z+$jets or $\gamma+$jets,
in which the jets are reconstructed as leptons.
In order to enhance the signal acceptance, looser selection criteria are
adopted for non-leading leptons.
The distribution of the observed events in the signal region is shown in
Fig. \ref{fig:ZZ_CDF_signal_region}, while in Table \ref{tab:ZZ_CDF_events}
is reported the comparison between the number of observed and expected events.
The production cross section for $ZZ$ events is evaluated according to:
\[
\sigma = \frac{N_{obs} - N_{bck}}{\mathcal{L} \cdot \epsilon \cdot \mathcal{B}}
\]
where
$\epsilon$ is the total efficiency,
$\mathcal{B}$ is the $ZZ \to l_a^+l_a^-l_b^+l_b^-$ branching fraction,
$N_{obs}$ the number of observed events,
$N_{bck}$ the expected number of background events and
$\mathcal{L}$ the total integrated luminosity.
The measured cross section is:
\[
\sigma(p \overline{p} \to ZZ) = 1.56^{+0.80}_{-0.63} \ (stat.) \pm 0.25 \ (syst.) \ \textnormal{pb}
\]
in good agreement with the previous measurement done by
CDF \cite{CDF_ZZ_First_Measurement} and D\O{} \cite{D0_ZZ_First_Observation}
collaborations, and with the theoretical expectation
$\sigma^{th}(p \overline{p} \to ZZ) = 1.4 \pm 0.1$ pb.

The statistical significance of the measurement is $5.70$ standard deviation, and
it is calculated evaluating the probability to have a number of events
equal or greater than than observed as a fluctuation
of the estimated background.

The analogous analysis performed by D\O{} collaboration on a data sample
corresponding to $1.7$ \mbox{fb$^{-1}$} \cite{D0_ZZ_First_Observation}, observes 3 events
with an expected background of $0.14^{+0.03}_{-0.02}$ events.
The significance of this measurement is 5.3 standard deviations.
The combination of this result with the measurement in the channel
$ZZ \to l^+l^-\nu\overline{\nu}$ increases the significance to 5.7 standard deviations
and measures a combined cross section
\[
\sigma(p \overline{p} \to ZZ) = 1.60 \pm 0.63 \ (stat.) \ ^{+0.16}_{-0.17} \ (syst.) \ \textnormal{pb}.
\]

\subsection{WV} 

\begin{figure}[!t]

  \begin{minipage}{.98\hsize}
   \begin{center}
     \includegraphics[width=.8\textwidth]{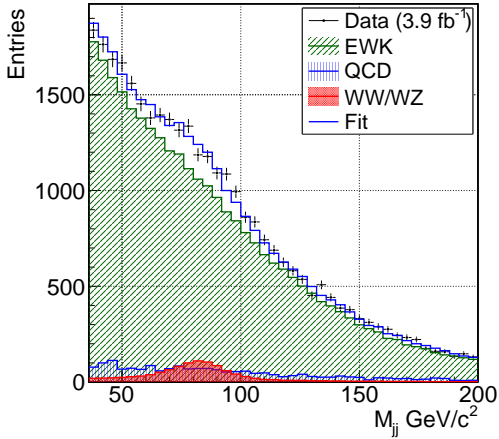}
     \includegraphics[width=.8\textwidth]{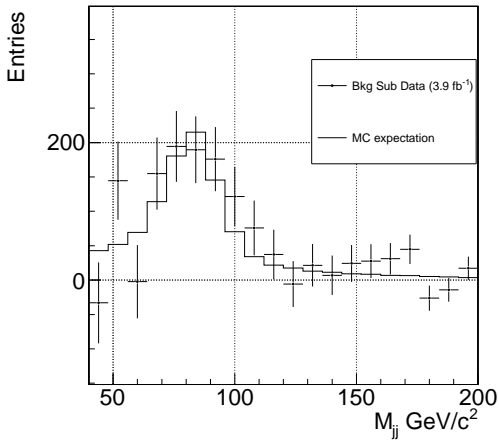}
   \end{center}
  \end{minipage}

  \caption{Invariant mass distribution for the hadronic boson candidate,
    compared with the signal and background distributions normalized
    according to the measured fractions (upper). Background-subtracted
    invariant mass distribution for the hadronic candidate in
    $WZ/WW \to l \nu jj$ events (lower).}

 \label{fig:WZ_WW_lnujj_CDF}
\end{figure}

The measurement of the production cross section of $WZ$ and $WW$ events
in $p \overline{p}$ collisions utilizing the decay channel $WV \to l \nu jj$
has been performed by the CDF collaboration on a data sample
corresponding to $3.9$ \mbox{fb$^{-1}$} \cite{CDF_WV_lnujj_mjj}.

In the sample of events acquired with high-$p_T$ electron or muon triggers,
candidate events are reconstructed requiring a well identified lepton ($e$ or $\mu$),
$\not \!\!\! E_T > 25$ GeV, consistent with the unobserved neutrino
from leptonic decay of the $W$ boson, and two jets with $E_T > 20$ GeV.
Furthermore the hadronic boson candidate is required to have an invariant mass
in the range ($30, 200$) \mbox{GeV/c$^2$} and $p_T > 40$ GeV/c.
The main processes contributing to the background are $W/Z+$jets, $t \overline{t}$ and
multi-jet events.
The fraction of diboson events is extracted by a binned fit to the invariant mass
distribution of the hadronic boson candidate.
The relative normalization of $W/Z$+jets and $t \overline{t}$ is set according to
their production cross sections, while a Gaussian constraint is applied to the
multi-jet component.
The expected value and width of the multi-jet constraint are extracted by a fit
to the $\not \!\!\! E_T$ distribution.
The invariant mass distribution of the hadronic boson candidate in the selected
events is shown in Fig. \ref{fig:WZ_WW_lnujj_CDF}
The measured cross section for the production of $WV$ events (with $V=Z, W$)
in $p \overline{p}$ collisions is:
\[
\sigma(p \overline{p} \to WV) = 14.4 \pm 3.1 \ (stat.) \pm 2.2 \ (syst.) \ \textnormal{pb}.
\]
The statistical significance of the measurement is $4.6$ standard deviations.

\begin{figure}[t]

  \begin{minipage}{\hsize}
   \begin{center}
     \includegraphics[width=.85\textwidth]{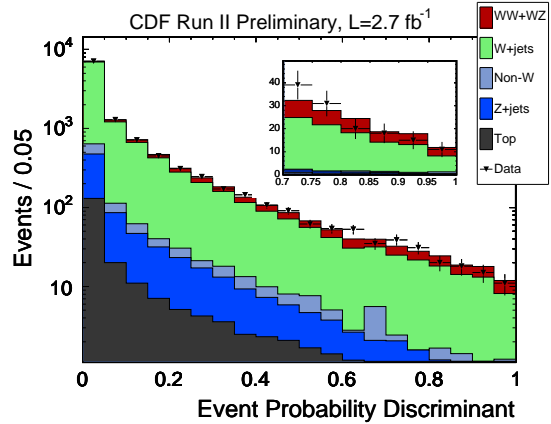}
   \end{center}
  \end{minipage}

  \caption{$EPD$ distribution for candidate diboson events compared with
    the likelihood fit of the sum of signal and background templates.}

 \label{fig:WZ_WW_CDF_EPD}
\end{figure}

The CDF collaboration analysed a subset of the data sample described above,
corresponding to an integrated luminosity of $2.7$ \mbox{fb$^{-1}$}, with 
a matrix element based technique, in order to maximize the use of collected
information for signal-background separation \cite{CDF_WV_lnujj_me}.

The procedure applied to calculate the event probability densities ($P_m$) under the
signal and background hypotheses is analogous to that described for the
$WW$ analysis.
The event probability densities are used to construct an Event Probability Discriminant
($EPD$), defined as:
\[
EPD = \frac{P_{WW} + P_{WZ}}{P_{WW} + P_{WZ} + \sum_i P_i}
\]
where the index $i$ runs over all the background sources.
The resulting template distributions for signal and background processes
are fit to the data using a binned likelihood approach (Fig. \ref{fig:WZ_WW_CDF_EPD}).
The measured cross section is:
\[
\sigma(p \overline{p} \to WV) = 17.7 \pm 3.1 \ (stat.) \pm 2.4 \ (syst.) \ \textnormal{pb}.
\]
The probability that the observed excess originated from a background fluctuation
(p-value) is $3.5 \cdot 10^{-8}$ which corresponds to a statistical significance of 5.4
standard deviations.

\subsection{VV $\to \not \! \! E_T + jj$ } 

\begin{figure}[b]

  \begin{minipage}{\hsize}
   \begin{center}
     \includegraphics[width=.75\textwidth]{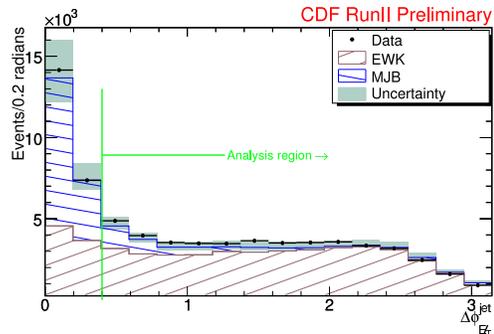}
   \end{center}
  \end{minipage}

  \caption{Distribution of the angle in the transverse plane between
    the $\not \!\!\! E_T$ and the closest jet with $E_T > 5$ GeV.
    The green line indicates the analysis region.}

 \label{fig:VV_MET_jj_dphi}
\end{figure}

The measurement of the production cross section of diboson events
in $p \overline{p}$ collisions utilizing hadronic final states,
has been performed by the CDF collaboration on a data sample
corresponding to $3.5$ \mbox{fb$^{-1}$} \cite{CDF_MET_jj}.

The events are reconstructed in final states with large $\not \!\!\! E_T$
and 2 identified jets, and are acquired with a set of $\not \!\!\! E_T$ based
triggers.
The analysed final state provide acceptance for $ZV$ events in which
the $Z$ boson decays in neutrinos and the other boson decays hadronically,
and also for $WV$ events in which the $W$ decays leptonically and the charged
lepton is not reconstructed.
The main processes contributing to the background are $W/Z+$jets and
multi-jet events.
In order to reduce the multijet background events with  $\not \!\!\! E_T^{sig} < 4$
or $\Delta \phi (\not \!\!\! E_T, j) < 0.4$ are rejected, where 
$\not \!\!\! E_T^{sig}$ is the $\not \!\!\! E_T$ significance and
$\Delta \phi (\not \!\!\! E_T, j)$ is the angle in the transverse plane between
the $\not \!\!\! E_T$ and the closest jet with $E_T > 5$ GeV.

\begin{figure}[!t]

  \begin{minipage}{\hsize}
   \begin{center}
     \includegraphics[width=.80\textwidth]{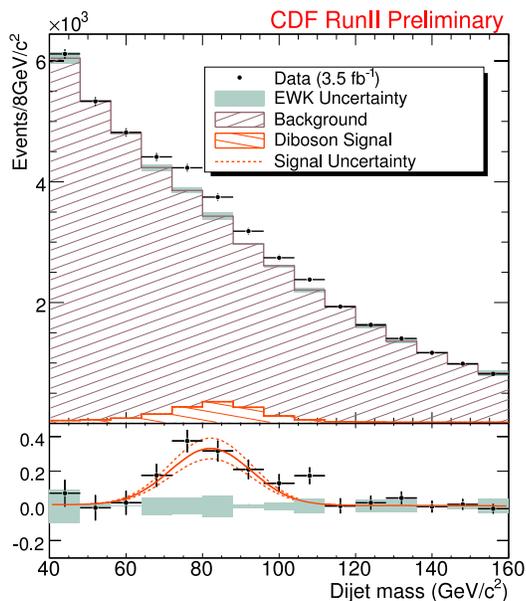}
   \end{center}
  \end{minipage}

  \caption{Distribution of the invariant mass of the hadronic boson candidate
    compared with signal and background distributions.
    The background-subtracted distribution is reported in the bottom plot.}

 \label{fig:VV_MET_jj}
\end{figure}

After all cuts are applied 44910 diboson candidate events are selected.
The final dijet mass fit is an unbinned extended maximum likelihood with
jet energy scale, and the slope and the normalization of the multijet background
treated as nuisance parameters and allowed to float in the fit within their
predetermined uncertainties.
The normalization of the $W/Z+$jets component is also freely floating
in the fit with no constraints.
The statistical significance of the measurement corresponds to $5.3$ standard deviations.

The measured cross section is:
\[
\sigma(p \overline{p} \to VV) = 18 \pm 2.8\ (stat.) \pm 2.4\ (syst.) \pm 1.1 \ (lumi.) \ \textnormal{pb}
\]
in agreement with the SM calculations of $16.8 \pm 0.5$ pb. 

In Fig. \ref{fig:VV_MET_jj} the selected candidate $VV$ events are compared with
the signal and background distributions, according to the fit results.

\section{Conclusions} 

Recent results on the study of the diboson production at Tevatron
have been presented.
Both the measurements of the production cross sections and the
limits on TGC parameters are competitive with the previous results from LEP
and consistent with the SM expectations.

\begin{figure}[t]

  \begin{minipage}{\hsize}
   \begin{center}
     \includegraphics[width=.90\textwidth]{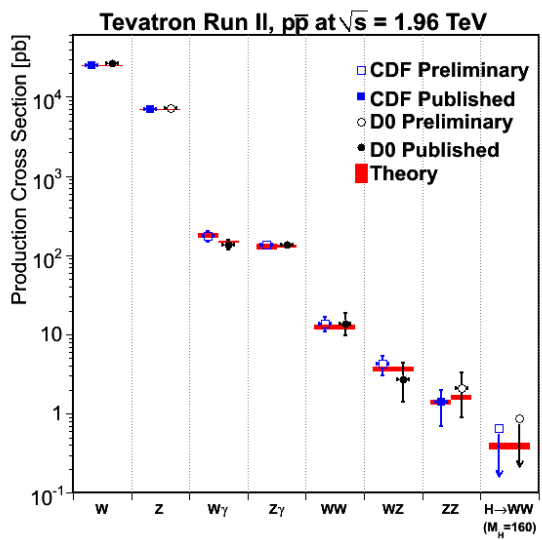}
   \end{center}
  \end{minipage}

  \caption{Summary of production cross sections measured by CDF
  and D\O{} collaborations for diboson processes, compared with
  SM expectations.}

 \label{fig:Xsec_comparison}
\end{figure}

The production cross sections measured for diboson processes
by CDF and D\O{} collaborations are summarized in Fig. \ref{fig:Xsec_comparison}

All the presented results will be further improved utilizing the increasing
available statistics of the collected data.

The Large Hadron Collider (LHC) will collect in the next years a data sample
of proton-proton collisions at a center-of-mass energy of $14$ TeV.
At this energy the production cross sections of diboson processes will
grow by approximately an order of magnitude, with respect to the Tevatron energy
\cite{ATLAS_expected performance}.
Since the design instantaneous luminosity of the LHC is
$\mathcal{L} = 10^{34}$ cm$^{-2}$ s$^{-1}$,
the diboson production rates will exceed those of the Tevatron by a factor
$\sim 100$.
Furthermore, because of the higher center-of-mass energy, the sensitivity of the
LHC experiments to the anomalous TGC's will be further enhanced.

\end{document}